\documentclass[aps,nofootinbib,showpacs,preprint]{revtex4}
\usepackage{graphicx}
\usepackage{amsfonts}

\begin{document}


\title{Giant Magnons and Spiky Strings on $S^3$ with $B$-field}

\author{Chiang-Mei Chen} \email{cmchen@phy.ncu.edu.tw}
\affiliation{Department of Physics and Center for Mathematics and Theoretical Physics, National Central University, Chungli 320, Taiwan}

\author{Jia-Hao Tsai} \email{immanuel.tsai@gmail.com}
\affiliation{Department of Physics, National Central University, Chungli 320, Taiwan}

\author{Wen-Yu Wen} \email{steve.wen@gmail.com}
\affiliation{Department of Physics, National Taiwan University, Taipei 106, Taiwan}

\date{\today}


\begin{abstract}
We study solutions for a rotating string on $S^3$ with a background NS-NS $B$-field and show the existence of spiky string and giant magnon as two limiting solutions.  We make a connection to the sine-Gordon model via the Polyakov worldsheet action and study the effect of $B$-field.  In particular, we find the magnon solution can be mapped to the excitation of a fractional spin chain. We conjecture a $B$-deformed SYM to be the gauge theory dual to this background.
\end{abstract}


\maketitle
\section{Introduction}
The AdS/CFT correspondence has revealed deep relation between string theory and gauge theory, in particular the correspondence between IIB strings on $AdS_5\times S^5$ and ${\cal N} = 4$ super Yang-Mills theory (SYM) at the t'Hooft large $N$ limit \cite{Maldacena:1997re, Gubser:1998bc, Witten:1998qj}. While the correspondence was mostly tested for those BPS states where partial supersymmetry protects it against quantum correction \cite{Aharony:1999ti}, semiclassical analysis on the near-BPS sector is also considered in \cite{Berenstein:2002jq}; the integrability of this Berenstein-Madalcena-Nastase (BMN) limit allows for quantitative tests of correspondence beyond BPS states, where the energy of classical string solutions is compared to the anomalous dimension of SYM operators with large R-charge. On the other hand, intimate relation between SYM dynamics and integrable spin chain was realized in \cite{Beisert:2003ea}, where the planar limit of the dilatation operator was identified with the Hamiltonian of integrable spin chains.  The correspondence between rotating string and spin chain was further explored in many papers, see \cite{Kruczenski:2003gt, Frolov:2003xy, Hernandez:2004uw, Bellucci:2004qr, Lunin:2005jy, Frolov:2005ty, Huang:2006bh, Wen:2006fw, Wen:2007az, Bak:2008vd} for a sample of it.  Recently, a giant magnon solution carrying momentum $p$ was first found on $S^2$ with dispersion relation \cite{Hofman:2006xt}:
\begin{equation}
E - J = \frac{\sqrt{\lambda}}{\pi} \left| \sin{\frac{p}{2}} \right|,
\end{equation}
where both energy $E$ and R-charge $J$ becomes infinite as $\lambda$ stays finite but large.  In addition, an infinitely winding solution with spiky configuration was also found with dispersion relation \cite{Ishizeki:2007we}:
\begin{equation}
E - T \Delta \phi = 2 T \bar{\theta_0},
\end{equation}
for angular difference between two spikes $\Delta \phi$ and spike height $\bar{\theta_0}$.  In fact, both solutions can be obtained from the same rotating string at different limits.  Since then many papers have been devoted to this study \cite{Chen:2006ge, Arutyunov:2006gs, Astolfi:2007uz, Minahan:2006bd, Chu:2006ae, Spradlin:2006wk, Kalousios:2006xy, Bobev:2006fg, Huang:2006vz, Bozhilov:2006bi, Kruczenski:2004wg, Kruczenski:2006pk, Ishizeki:2007we, Mosaffa:2007ty, Maldacena:2006rv, Plefka:2005bk, Lee:2008sk, David:2008yk, Bozhilov:2007km, Dimov:2007ey, Kluson:2007fr, Ishizeki:2007kh, Kluson:2007st, Kluson:2008wv, Abbott:2008yp, Ahn:2008sk, Lee:2008ui, Ahn:2008hj, Gomez:2006va}.  In particular, solutions with nontrivial NS-NS $B$-field were also investigated in various context, such as in the Melvin twisted background \cite{Huang:2006vz, Lee:2008sk}, Lunin-Maldacena background \cite{Chu:2006ae}, and $S^2$ with NS-NS $B$-field \cite{Lee:2008sk}.  In this paper, we consider a rotating string on $S^3$ with background NS-NS $B$-field.  This background can be obtained in the type IIB strings with appropriate embedding ansatz.  We obtain both solutions of spiky string and giant magnon as two limiting cases, providing appropriate choice of integral constants.  As an integrable system, we find it is straightforward to write down the corresponding sine-Gordon model upon a redefinition of mass of sine-Gordon field, and the scattering and bound state of magnons are ready to be translated into solutions of multi-kinks.  In the spirit of the AdS/CFT correspondence, we also expect the existence of a dual description given by a SYM deformed by such $B$-field.  This conjecture is later supported by providing a dictionary established between dispersion relation of magnon solution in the gravity side and excitation of spin chain formed by the long trace operator in the proposed $B$-deformed SYM.

The outline of this paper is as follows: first in the section II and III, we present a rotating string solution on different coordinate systems of $S^3$ with NS-NS $B$-field using the Nambu-Goto string action.  In the two limiting cases, we obtain the solution for spiky string and giant magnon.  In the section IV, we take advantage of the Polyakov action in order to make a connection to the sine-Gordon model.  In the section V, we investigate the obtained solutions within the description of spin chain, and conjecture the existence of a $B$-deformed super Yang-Mills theory, which is dual to our string background with NS-NS $B$-field.  At last, we have a discussion in the section VI.

\section{String on $S^3$ with $B$-field}
We consider a rigid string rotating on $S^3$ with background NS-NS $B$-field.  The relevant metric reads,
\begin{equation}
ds^2 = - dt^2 + d\vartheta^2 + \sin^2\vartheta \, d\varphi_1^2 + \cos^2\vartheta \, d\varphi_2^2,
\end{equation}
with magnetic NS-NS $B$-field
\begin{equation}
B = b \, \sin^2\vartheta \, d\varphi_1 \wedge d\varphi_2.
\end{equation}
This background can be obtained from the type IIB strings with appropriate embedding ansatz.  One possible construction is to consider a near-horizon geometry of intersecting NS1 and NS5 branes \cite{Horowitz:1991cd, Gauntlett:1997cv}.  The magnetic component of NS-NS flux supported by NS5 branes fills the $S^3$ with constant field strength, while the electric component is decoupled from the string worldsheet.

The corresponding Nambu-Goto (NG) Lagrangian is
\begin{eqnarray}
{\cal L}_{\rm NG}^{S^3} &=& - T \Biggl\{ \Bigl[ (- \partial_\tau t \partial_\sigma t + \partial_\tau \vartheta \partial_\sigma \vartheta + \sin^2\vartheta \partial_\tau \varphi_1 \partial_\sigma \varphi_1 + \cos^2\vartheta \partial_\tau \varphi_2 \partial_\sigma \varphi_2)^2
\nonumber\\
&& \qquad - \left( - (\partial_\tau t)^2 + (\partial_\tau \vartheta)^2 + \sin^2\vartheta (\partial_\tau \varphi_1)^2 + \cos^2\vartheta (\partial_\tau \varphi_2)^2 \right) \cdot
\nonumber\\
&& \qquad\quad \left( - (\partial_\sigma t)^2 + (\partial_\sigma \vartheta)^2 + \sin^2\vartheta (\partial_\sigma \varphi_1)^2 + \cos^2\vartheta (\partial_\sigma \varphi_2)^2 \right) \Bigr]^{\frac12}
\nonumber\\
&& \qquad - b \sin^2\vartheta (\partial_\tau \varphi_1 \partial_\sigma \varphi_2 - \partial_\sigma \varphi_1 \partial_\tau \varphi_2) \Biggr\},
\end{eqnarray}
where $T = \sqrt{\lambda}/2 \pi$ is the string tension.  We assume the string rotates in $\varphi_1$ and $\varphi_2$ directions, with the following embedding ansatz:
\begin{equation}
t = \kappa \tau, \qquad \vartheta = \vartheta(\sigma), \qquad \varphi_1 = \nu_1 \tau + \sigma, \qquad \varphi_2 = \nu_2 \tau + \varphi(\sigma).
\end{equation}
The solution of $\vartheta'$, in terms of $\varphi'$, can be obtained by solving the equation for $t$:
\begin{equation}
\vartheta'^2 = \frac{(\kappa^2 - C_1^2) (\nu_1 \sin^2\vartheta + \nu_2 \cos^2\vartheta \; \varphi')^2}{C_1^2 (\kappa^2 - \nu_1^2 \sin^2\vartheta - \nu_2^2 \cos^2\vartheta)} - \sin^2\vartheta - \cos^2\vartheta \; \varphi'^2,
\end{equation}
and the $\varphi'$ can be solved by the ratio of equations for $t$ and $\varphi_1$:
\begin{equation}
\varphi' = \frac{\sin^2\vartheta \left[ \kappa (\kappa C_1 - \nu_1 C_2) - \nu_2^2 C_1 \cos^2\vartheta + b \kappa \nu_1 \nu_2 \sin^2\vartheta \right]}{\nu_2 \cos^2\vartheta (\kappa C_2 - \nu_1 C_1 \sin^2\vartheta - b \kappa \nu_2 \sin^2\vartheta)}.
\end{equation}
The general expression of $\vartheta'$ is rather complicated, to be explicit as follows:
\begin{eqnarray}
\vartheta' &=& - \frac{\kappa \sin\vartheta \sqrt{ \alpha_6 \cos^6\vartheta + \alpha_4 \cos^4\vartheta + \alpha_2 \cos^2\vartheta + \alpha_0}}{\nu_2 \cos\vartheta (\kappa C_2 - C_1 \nu_1 \sin^2\vartheta - b \kappa \nu_2 \sin^2\vartheta)},
\\
\alpha_6 &=& - \nu_2^2 (\nu_1^2 - \nu_2^2) (1 - b^2),
\nonumber\\
\alpha_4 &=& - \nu_2 \left[ \nu_2 (C_1^2 - 2 \nu_1^2 + \nu_2^2 + \kappa^2) + 2 b (C_2 \nu_2^2 - C_2 \nu_1^2 + C_1 \kappa \nu_1) + b^2 \nu_2 (3 \nu_1^2 - 2 \nu_2^2) \right],
\nonumber\\
\alpha_2 &=& - \alpha_0 + \nu_2 \left[ \nu_2 (C_1^2 - C_2^2 - \nu_1^2 + \kappa^2) + 2 b (C_2 \nu_2^2 - C_2 \nu_1^2 + C_1 \kappa \nu_1) + b^2 \nu_2 (2 \nu_1^2 - \nu_2^2) \right],
\nonumber\\
\alpha_0 &=& - (C_1 \kappa - C_2 \nu_1 + b \nu_1 \nu_2)^2.
\nonumber
\end{eqnarray}
The desired boundary condition $\vartheta' = 0$ at $\vartheta = \pi/2$ requires $\alpha_0 = 0, \alpha_2 = 0$ which give two possible constraints:
\begin{equation}
C_1 = \nu_1, \quad C_2 = \kappa + b \nu_2; \qquad \kappa = \nu_1, \quad C_2 = C_1 + b \nu_2;
\end{equation}
corresponding to spiky string and giant magnon respectively.

\subsection{Spiky string}
The first choice of parameters effectively imposes the following conditions at $\vartheta \to \pi/2$: (i) $\varphi'$ is regular by $C_2 = \kappa + b \nu_2$ and (ii) $\vartheta' \to 0$ by $C_1 = \nu_1$. Then the solutions of $\varphi'$ and $\vartheta'$ reduce to
\begin{eqnarray}
\varphi' &=& \frac{\nu_1 (\nu_2 + b \kappa) \sin^2\vartheta}{\nu_1^2 \sin^2\vartheta - b \kappa \nu_2 \cos^2\vartheta - \kappa^2},
\\
\vartheta' &=& \frac{\kappa \sqrt{(\nu_1^2 - \nu_2^2) (1 - b^2)} \; \sin\vartheta \cos\vartheta}{\nu_1^2 \sin^2\vartheta - b \kappa \nu_2 \cos^2\vartheta - \kappa^2} \sqrt{\sin^2\vartheta - \sin^2\vartheta_0},\label{eqn:vartheta}
\end{eqnarray}
where
\begin{equation}
\sin\vartheta_0 = \frac{\kappa + b \nu_2}{\sqrt{(\nu_1^2 - \nu_2^2)(1 - b^2)}}.
\end{equation}

\begin{figure}
\includegraphics[width=0.45\textwidth]{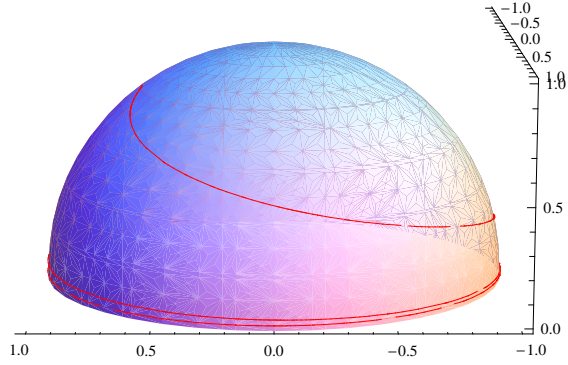}
\caption{Snapshots of spiky string on $S^2$ section $ds^2 = d\vartheta^2 + \sin^2{\vartheta} d\varphi_1^2$ with $b = 0$.  It shows the string wraps around the equator an infinite number of times.}
\label{fig:spike}
\end{figure}

The conserved quantities, energy and two angular momenta, are given by
\begin{eqnarray}
E &=&
- 2 \int_{\vartheta_0}^{\frac{\pi}2} \, \frac{d\vartheta}{\vartheta'} \, \frac{\partial {\cal L}_{\rm NG}^{S^3}}{\partial \dot t}
= 2 T \frac{(\nu_1^2 - \kappa^2) \ln\frac{2\cos\vartheta_0}{\cos\frac{\pi}2}}{\kappa \sqrt{(\nu_1^2 - \nu_2^2) (1 - b^2)} \; \cos\vartheta_0},
\label{ESpikyS3} \\
J_1 &=& 2 \int_{\vartheta_0}^{\frac{\pi}2} \, \frac{d\vartheta}{\vartheta'} \, \frac{\partial {\cal L}_{\rm NG}^{S^3}}{\partial \dot \varphi_1}
= 2 T \left[ \frac{\nu_1 (1 - b^2) \cos\vartheta_0}{\sqrt{(\nu_1^2 - \nu_2^2)(1 - b^2)}} + \frac{b \nu_1 (\nu_2 + b \kappa) \ln\frac{2 \cos\vartheta_0}{\cos\frac{\pi}2}}{\kappa \sqrt{(\nu_1^2 - \nu_2^2) (1 - b^2)} \cos\vartheta_0} \right],
\label{J1SpikyS3} \\
J_2 &=& 2 \int_{\vartheta_0}^{\frac{\pi}2} \, \frac{d\vartheta}{\vartheta'} \, \frac{\partial {\cal L}_{\rm NG}^{S^3}}{\partial \dot \varphi_2}
= - 2 T \left[ \frac{\nu_2 (1 - b^2) \cos\vartheta_0}{\sqrt{(\nu_1^2 - \nu_2^2)(1 - b^2)}} + \frac{b (\nu_1^2 - \kappa^2) \ln\frac{2 \cos\vartheta_0}{\cos\frac{\pi}2}}{\kappa \sqrt{(\nu_1^2 - \nu_2^2) (1 - b^2)} \cos\vartheta_0} \right]. \label{J2SpikyS3}
\end{eqnarray}
The terms with $\ln\frac{2 \cos\vartheta_0}{\cos\frac{\pi}2}$ represent the divergent part of the physical quantities, indicating the energy diverges in general.   The longitude angle between two spikes,
\begin{eqnarray}
\Delta \phi = 2 \int_{\vartheta_0}^{\frac{\pi}2} \frac{d\vartheta}{\vartheta'}
= - 2 \left( \frac{\pi}2 - \vartheta_0 \right) + \frac{2 (\nu_1^2 - \kappa^2) \ln\frac{2 \cos\vartheta_0}{\cos\frac{\pi}2}}{\kappa \sqrt{(\nu_1^2 - \nu_2^2) (1 - b^2)} \cos\vartheta_0},
\end{eqnarray}
is also a divergent quantity.  Their combination, however, renders a finite value:
\begin{equation}
\frac1{2T} (E - T \Delta \phi) = \frac{\pi}2 - \vartheta_0 = \frac{p}2.
\end{equation}
The spiky string can be visualized as a string wrapped around the equator an infinite number of times, as shown in the Fig. \ref{fig:spike}.  The divergent quantities $E$ and $\Delta\phi$ are due to the infinite length of string.  However, the latitude angle from peak of spike to the north pole is nothing but $\vartheta_0$, which stays finite.  This relation holds independently from the $B$-field.

Moreover, the angular momenta of string, $J_1$ and $J_2$, are also divergent in general, except for two special values of parameters. In the first case, $B$-field is absent, i.e. $b = 0$, and we have the dispersion relation analog with that of single spiky string \cite{Ishizeki:2007we}
\begin{equation}
J_1 = \sqrt{J_2^2 + 4 T^2 \cos^2\vartheta_0} = \sqrt{J_2^2 + \frac{\lambda}{\pi^2} \sin^2\frac{p}2}.
\end{equation}
However, in this analog, $J_1$ is related to the energy of spiky string and the momentum of spiky string can also be identified as $p = \pi - 2 \vartheta_0$.  In the other case, we have $\nu_1^2 = \kappa^2$ and  $\nu_2 = - b \kappa$, which, however, reduces to a trivial configuration, namely $\vartheta_0 = \pi/2$ and $J_1 = J_2 = 0$.

In general, one can use $E$ to {\it regularize} $J_1$ and $J_2$ as follows:
\begin{eqnarray}
\tilde J_1 &=& J_1 - \frac{b \nu_1 (\nu_2 + b \kappa)}{\nu_1^2 - \kappa^2} E = 2 T \frac{\nu_1 (1 - b^2) \cos\vartheta_0}{\sqrt{(\nu_1^2 - \nu_2^2)(1 - b^2)}},
\\
\tilde J_2 &=& J_2 + b E = - 2 T \frac{\nu_2 (1 - b^2) \cos\vartheta_0}{\sqrt{(\nu_1^2 - \nu_2^2)(1 - b^2)}},
\end{eqnarray}
and to derive the dispersion relation:
\begin{equation}
\tilde J_1 = \sqrt{\tilde J_2^2 + 4 T^2 (1 - b^2) \cos^2\vartheta_0}.
\end{equation}

In order to picture the explicit profile of spiky string, we can integrate (\ref{eqn:vartheta}) and obtain a relation between $\sigma$ and $\vartheta$
\begin{equation}
\pm\kappa \sqrt{(\nu_1^2 - \nu_2^2) (1 - b^2)} \; \sigma = - \frac{\kappa^2 - \nu_1^2}{\cos\vartheta_0} \mathrm{arccosh}{\frac{\cos\vartheta_0}{\cos\vartheta}} + \frac{\kappa (\kappa + b \nu_2)}{\sin\vartheta_0} \arccos{\frac{\sin\vartheta_0}{\sin\vartheta}},
\end{equation}
where there is no bound for $\sigma$.  In fact, one can easily see that $\sigma\to \pm\infty$ as $\vartheta\to \frac{\pi}{2}_-$. The relation for $\varphi$ is complicated but can be obtained by iteration.



\subsection{Giant magnon}
The other choice of parameter, corresponding to giant magnon, is for $\kappa = \nu_1$ and $C_2 = C_1 + b \nu_2$.  We then obtain
\begin{eqnarray}
\varphi' &=& - \frac{(C_1 \nu_2 + b \nu_1^2) \sin^2\vartheta}{\nu_1 (C_1 + b \nu_2) \cos^2\vartheta},\label{mag:varphi}
\\
\vartheta' &=& \frac{\sin\vartheta}{\sin\vartheta_1 \, \cos\vartheta} \sqrt{\sin^2\vartheta - \sin^2\vartheta_1},\label{mag:vartheta}
\end{eqnarray}
where
\begin{equation}
\sin\vartheta_1 = \frac{C_1 + b \nu_2}{\sqrt{(\nu_1^2 - \nu_2^2)(1 - b^2)}}.
\end{equation}

\begin{figure}
\includegraphics[width=0.45\textwidth]{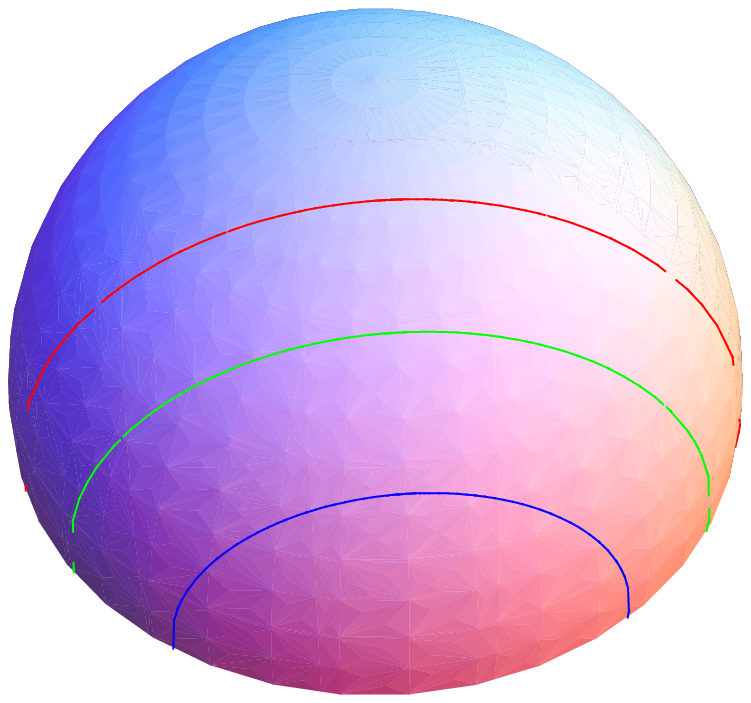}
\includegraphics[width=0.45\textwidth]{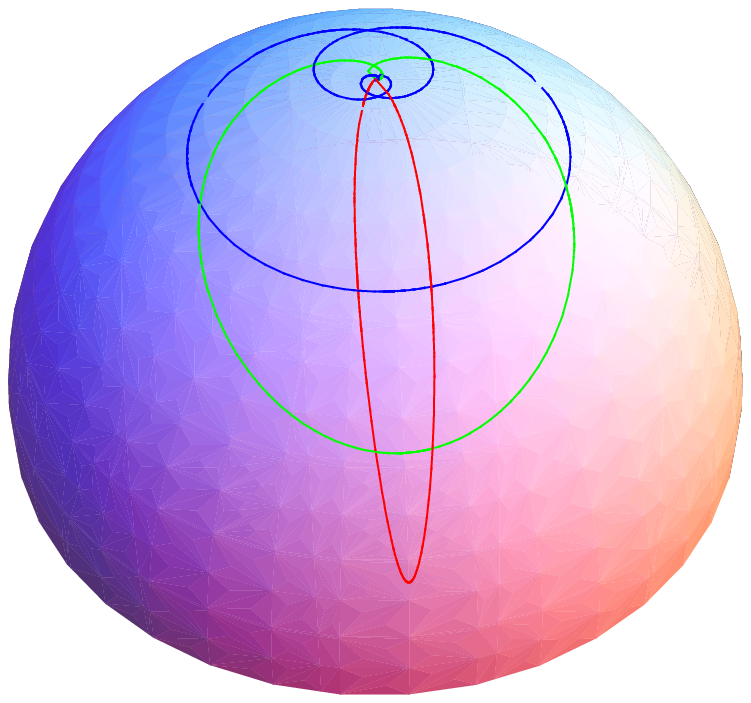}
\caption{Snapshots of giant magnon on $S^2$ section of $ds^2 = d\vartheta^2 + \sin^2{\vartheta} d\varphi_1^2 + \cos^2{\vartheta} d\varphi_2^2$ with various $B$-field, where $b = 0$ (red), $0.5$ (green), $0.7$ (blue).  To the left, string configuration in $(\vartheta, \varphi_1)$.  To the right, string configuration in $(\vartheta, \varphi_2)$.}
\label{fig:magnon}
\end{figure}

The conserved quantities, energy and two angular momenta, are given as follows:
\begin{eqnarray}
E &=& - 2 \int_{\vartheta_1}^{\frac{\pi}2} \, \frac{d\vartheta}{\vartheta'} \, \frac{\partial {\cal L}_{\rm NG}}{\partial \dot t}
= 2 T \frac{(\nu_1^2 - C_1^2) \ln\frac{2 \cos\vartheta_1}{\cos\frac{\pi}2}}{\nu_1 \sqrt{(\nu_1^2 - \nu_2^2) (1 - b^2)} \cos\vartheta_1},
\label{EMagnonS3} \\
J_1 &=& \! 2 \int_{\vartheta_1}^{\frac{\pi}2} \, \frac{d\vartheta}{\vartheta'} \, \frac{\partial {\cal L}_{\rm NG}}{\partial \dot \varphi_1}
\!=\! 2 T \left[ \frac{- \nu_1 (1 - b^2) \cos\vartheta_1}{\sqrt{(\nu_1^2 - \nu_2^2)(1 - b^2)}} \!+\! \frac{(\nu_1^2 \!-\!  C_1^2 \!-\! C_1 b \nu_2 \!-\! b^2 \nu_1^2) \ln\frac{2 \cos\vartheta_1}{\cos\frac{\pi}2}}{\nu_1 \sqrt{(\nu_1^2 - \nu_2^2) (1 - b^2)} \cos\vartheta_1} \right],
\label{J1MagnonS3} \\
J_2 &=& 2 \int_{\vartheta_1}^{\frac{\pi}2} \, \frac{d\vartheta}{\vartheta'} \, \frac{\partial {\cal L}_{\rm NG}}{\partial \dot \varphi_2}
= 2 T \frac{\nu_2 (1 - b^2) \cos\vartheta_1}{\sqrt{(\nu_1^2 - \nu_2^2)(1 - b^2)}}. \label{J2MagnonS3}
\end{eqnarray}
We have string configuration plotted in the Fig. \ref{fig:magnon} for various values of $B$-field.  The longitude angle spanned between two ends of string is
\begin{equation}
\Delta \phi = 2 \int_{\vartheta_1}^{\frac{\pi}2} \frac{d\vartheta}{\vartheta'}
= \pi - 2 \vartheta_1 = p,
\end{equation}
where the momentum of magnon is identified as $p = \pi - 2 \vartheta_1$. This relation again holds independently of the $B$-field. The energy, $E$, and one angular momentum, $J_1$, are divergent. However, we can rescale energy as
\begin{equation}
\tilde E = \frac{\nu_1^2 - C_1^2 - C_1 b \nu_2 - b^2 \nu_1^2}{\nu_1^2 - C_1^2} E = \left( 1 - b \frac{C_1 \nu_2 + b \nu_1^2}{\nu_1^2 - C_1^2} \right) E,
\end{equation}
and then obtain the following dispersion relation:
\begin{equation}
\tilde E -  J_1 = \sqrt{J_2^2 + 4 T^2 (1 - b^2) \cos^2\vartheta_1} = \sqrt{J_2^2 + \frac{\lambda}{\pi^2} (1 - b^2) \sin^2\frac{p}2}.
\end{equation}

Similarly, for the explicit profile of giant magnon we can integrate (\ref{mag:varphi}) and (\ref{mag:vartheta}),
then we have simple relations:
\begin{equation}
\varphi = - \frac{(C_1 \nu_2 + b \nu_1^2) \; \mathrm{arctanh}(\tan\vartheta_1 \tan\sigma)}{\nu_1 \sqrt{(\nu_1^2 - \nu_2^2)(1 - b^2)} \, \cos\vartheta_1}, \qquad \sin\vartheta = \frac{\sin\vartheta_1}{\cos\sigma}.
\end{equation}
This implies the finite range of $\sigma$, i.e. $|\sigma| \le \pi/2 - \vartheta_1$.

\section{String on $S^1 \times S^2$ with $B$-field}
In this section, we study a rotating string in a different coordinate system.  Let us transform $S^3$ to a new angular coordinate system:
\begin{equation}\label{hopf}
\vartheta = \frac12 \theta, \qquad \varphi_1 = \frac12 (\phi - \psi), \qquad \varphi_2 = \frac12 (\phi + \psi),
\end{equation}
such that the $S^3$ metric is written as the Hopf fibration 3-sphere\footnote{The ranges of parameters are: $0 \le \vartheta \le \pi, \; 0 \le \varphi_1, \varphi_2 \le 2 \pi; \quad 0 \le \theta, \phi \le 2 \pi, \; 0 \le \psi \le 4 \pi.$}, i.e. $S^3 \sim S^1 \times S^2$
\begin{equation}
ds^2 = - dt^2 + \frac14 \left[ d\theta^2 + \sin^2\theta d\phi^2 + (d\psi + \cos\theta d\phi)^2 \right],
\end{equation}
and the B field becomes
\begin{equation}
B = \frac{b}4 (1 - \cos\theta) d\phi \wedge d\psi.
\end{equation}
Straightforwardly, the solution is transformed as follows:
\begin{equation}
t = \kappa \tau, \qquad \theta = 2 \vartheta(\sigma), \qquad \phi = (\nu_2 + \nu_1) \tau + \varphi(\sigma) + \sigma, \quad \psi = (\nu_2 - \nu_1) \tau + \varphi(\sigma) - \sigma.
\end{equation}

Instead of repeating the same calculation, we now consider a more general background
\begin{eqnarray}
ds^2 &=& - dt^2 + k^2 \left[ d\theta^2 + \sin^2\theta d\phi^2 + (d\psi + \cos\theta d\phi)^2 \right],
\nonumber\\
B &=& (b_0 + b_1 \cos\theta + b_2 \cos\phi) \, d\phi \wedge d\psi,
\end{eqnarray}
which recovers the $S^3$ case for the special values $k^2 = 1/4, b_0 = - b_1 = b/4, b_2 = 0$. Then the NG action becomes
\begin{eqnarray}
{\cal L}_{\rm NG}^{S^1\times S^2} &=& - T \Biggl\{ \Bigl[ \left( - \partial_\tau t \partial_\sigma t + k^2 \partial_\tau \theta \partial_\sigma \theta + k^2 \sin^2\theta \partial_\tau \phi \partial_\sigma \phi + k^2 (\partial_\tau \psi + \cos\theta \partial_\tau \phi) (\partial_\sigma \psi + \cos\theta \partial_\sigma \phi) \right)^2
\nonumber\\
&& \qquad - \left( - (\partial_\tau t)^2 + k^2 (\partial_\tau \theta)^2 + k^2 \sin^2\theta (\partial_\tau \phi)^2 + k^2 (\partial_\tau \psi + \cos\theta \partial_\tau \phi)^2 \right) \cdot
\nonumber\\
&& \qquad\quad \left( - (\partial_\sigma t)^2 + k^2 (\partial_\sigma \theta)^2 + k^2 \sin^2\theta (\partial_\sigma \phi)^2 + k^2 ( \partial_\sigma \psi + \cos\theta \partial_\sigma \phi)^2 \right) \Bigr]^{\frac12}
\nonumber\\
&& \qquad - (b_0 + b_1 \cos\theta + b_2 \cos\phi) (\partial_\tau \phi \partial_\sigma \psi - \partial_\tau \psi \partial_\sigma \phi) \Biggr\}.
\end{eqnarray}

Assuming a typical form of solution
\begin{equation}
t = \kappa \tau, \qquad \theta = \theta(\sigma), \qquad \phi = \mu_1 \tau + \sigma, \qquad \psi = \mu_2 \tau + \psi(\sigma),
\end{equation}
the general solution for $\psi'$ is\footnote{The parameters $b_0, b_2$ do not appear in the solution since they effectively are total derivative terms in the NG action.}
\begin{equation}
\psi' = - \frac{C_1 (\kappa^2 - k^2 \mu_2^2 \sin^2\theta) - \kappa (C_2 - b_1 \mu_2 \cos\theta) (\mu_1 + \mu_2 \cos\theta)}{C_1 (\kappa^2 \cos\theta + k^2 \mu_1 \mu_2 \sin^2\theta) - \kappa (C_2 - b_1 \mu_2 \cos\theta) (\mu_2 + \mu_1 \cos\theta)},
\end{equation}
which is solved from the equation of $\phi$.\footnote{There is another version of solution by solving equation of $\psi$:
\begin{equation}
\psi' = - \frac{C_1 (\kappa^2 \cos\theta + k^2 \mu_1 \mu_2 \sin^2\theta) - \kappa (C_2 + b_1 \mu_1 \cos\theta) (\mu_1 + \mu_2 \cos\theta)}{C_1 (\kappa^2 - k^2 \mu_1^2 \sin^2\theta) - \kappa (C_2 + b_1 \mu_1 \cos\theta) (\mu_2 + \mu_1 \cos\theta)},
\end{equation}
which is inverse of the solution from equation of $\phi$ together with changing parameters $(\mu_1, \mu_2, b_1) \to (\mu_2, \mu_1, -b_1)$. After imposing $C_1 \kappa = (\mu_2 - \mu_1) (C_2 - b_1 \mu_1)$,
one obtains the desired $\psi'$.  This duality can be understood as exchange of coordinates $\phi$ and $\psi$, while $B$-field has to flip the sign simultaneously.
}
The expression of $\theta'$ is
\begin{eqnarray}
\theta' &=& - \frac{\kappa \sin\theta \sqrt{\alpha_3 \cos^3\theta + \alpha_2 \cos^2\theta + \alpha_1 \cos\theta + \alpha_0}}{C_1 (\kappa^2 \cos\theta + k^2 \mu_1 \mu_2 \sin^2\theta) - \kappa (C_2 - b_1 \mu_2 \cos\theta) (\mu_2 + \mu_1 \cos\theta)},
\\
\alpha_3 &=& 2 \mu_1 \mu_2^3 (k^4 - b_1^2),
\nonumber\\
\alpha_2 &=& \mu_2^2 \left[ (k^4 - b_1^2) (\mu_1^2 + \mu_2^2) - k^2 (C_1^2 + \kappa^2) - 2 b_1 (C_1 \kappa - 2 C_2 \mu_1) \right],
\nonumber\\
\alpha_1 &=& - 2 \mu_2 \left[ C_2^2 \mu_1 - C_1 C_2 \kappa + k^4 \mu_1 \mu_2^2 + b_1 (C_1 \kappa \mu_1 - C_2 \mu_1^2 - C_2 \mu_2^2) \right],
\nonumber\\
\alpha_0 &=& - (C_1 \kappa - C_2 \mu_1)^2 - \mu_2^2 \left[ (\mu_1^2 + \mu_2^2) k^4 - (C_1^2 + \kappa^2) k^2 + C_2^2 \right].
\nonumber
\end{eqnarray}
Let us analyze the boundary conditions at the equator; at $\theta = \pi$, we generally have a {\it fixed} values for $\psi' = 1$ and $\theta' = 0$, regardless of the parameters.  In order to relax this restriction, we should impose
\begin{equation} \label{S1S2C1}
C_1 \kappa = (\mu_1 - \mu_2) (C_2 + b_1 \mu_2),
\end{equation}
and then $\psi'$ and $\theta'$ become
\begin{eqnarray}
\theta' &=& \frac{\kappa \sin\theta \sqrt{\alpha_3 \cos^3\theta + \alpha_2 \cos^2\theta + \alpha_1 \cos\theta + \alpha_0}}{\mu_1 (1 + \cos\theta) [C_2 \kappa - C_1 k^2 \mu_2 + \mu_2 ( C_1 k^2 - b_1 \kappa) \cos\theta]},
\\
\psi' &=& - \frac{\kappa \mu_2 (1 + \cos\theta) [C_2 - b_1 \mu_1 + b_1 \mu_2 (1 - \cos\theta)] + C_1 k^2 \mu_2^2 \sin^2\theta}{\mu_1 (1 + \cos\theta) [C_2 \kappa - C_1 k^2 \mu_2 + \mu_2 ( C_1 k^2 - b_1 \kappa) \cos\theta]}.
\end{eqnarray}
Now we can avoid $\theta'$ to be vanishing at the equator if one can factor out $(1 + \cos\theta)$ from the square root.  This requires
\begin{equation}
\alpha_1 - 2 \alpha_2 + 3 \alpha_3 = 0, \qquad \alpha_0 - \alpha_2 + 2 \alpha_3 = 0,
\end{equation}
which give two possible constraints
\begin{equation}
C_1 = k (\mu_1 - \mu_2); \qquad C_1 k = C_2 + b_1 \mu_2.
\end{equation}
Together with the constraint (\ref{S1S2C1}), we arrive at two choices of parameters corresponding to spiky string and giant magnon respectively:
\begin{equation}\label{S1S2Constraint}
C_1 = k (\mu_1 - \mu_2), \quad C_2 = \kappa k - b_1 \mu_2; \qquad \kappa = k (\mu_1 - \mu_2), \quad C_2 = C_1 k - b_1 \mu_2.
\end{equation}

\subsection{Spiky string}
The spiky string is given by the first choice in (\ref{S1S2Constraint}), $C_1 = k (\mu_1 - \mu_2)$ and $C_2 = \kappa k - b_1 \mu_2$, and the $\theta'$ and $\psi'$ are simplified as
\begin{eqnarray}
\theta' &=& \frac{\kappa \sqrt{2 \mu_1 \mu_2 (b_1^2 - k^4)} \sin\theta \sqrt{\cos\theta_0 - \cos\theta}}{\mu_1 (k^3 \mu_1 - k^3 \mu_2 - b_1 \kappa) \cos\theta + k \kappa^2 - k^3 \mu_1^2 + k^3 \mu_1 \mu_2 - b_1 \kappa \mu_2},
\\
\psi' &=& \frac{\mu_2 (k^3 \mu_1 - k^3 \mu_2 + b_1 \kappa) \cos\theta - k \kappa^2 + k^3 \mu_2^2 - k^3 \mu_1 \mu_2 + b_1 \kappa \mu_1}{\mu_1 (k^3 \mu_1 - k^3 \mu_2 - b_1 \kappa) \cos\theta + k \kappa^2 - k^3 \mu_1^2 + k^3 \mu_1 \mu_2 - b_1 \kappa \mu_2},
\end{eqnarray}
where
\begin{equation}
\cos\theta_0 = 1 - \frac{(k \kappa - b_1 \mu_1 - b_1 \mu_2)^2}{2 \mu_1 \mu_2 (b_1^2 - k^4)}.
\end{equation}
The energy of spiky string is
\begin{eqnarray}
E &=& - 2 \int_{\theta_0}^\pi \, \frac{d\theta}{\theta'} \, \frac{\partial {\cal L}_{\rm NG}^{S^1\times S^2}}{\partial \dot t}
= 2 T \frac{\sqrt2 k^2 [\kappa^2 - k^2 (\mu_1 - \mu_2)^2)]}{\kappa \sqrt{2 \mu_1 \mu_2 (b_1^2 - k^4)} \cos\frac{\theta_0}2} \ln \frac{2 \cos\frac{\theta_0}2}{\cos\frac{\pi}2}.
\end{eqnarray}
However, for the particular values: $b_1 = -b/4, k^2 = 1/4, \mu_1 = \nu_2 + \nu_1, \mu_2 = \nu_2 - \nu_1$, then we have
\begin{eqnarray}
&& \sin\frac{\theta_0}2 = \sqrt{\frac{1 - \cos\theta_0}2} = \frac{\kappa + b \nu_2}{\sqrt{(\nu_1^2 - \nu_2^2)(1 - b^2)}},
\nonumber\\
&& E = \frac{2 T (\kappa^2 - \nu_1^2)}{\kappa \sqrt{(\nu_1^2 - \nu_2^2) (1 - b^2)} \cos\frac{\theta_0}2} \ln\frac{2 \cos\frac{\theta_0}2}{\cos\frac{\pi}2},
\end{eqnarray}
which is consistent with (\ref{ESpikyS3}) with opposite sign due to the different branch solution of $\theta'$. 

Unlike the energy, the angular momenta do include contribution from $b_0$ and $b_2$, therefore be more complicated. Their expressions are
\begin{eqnarray}
J_1 &=& 2 \int_{\theta_0}^\pi \, \frac{d\theta}{\theta'} \, \frac{\partial {\cal L}_{\rm NG}^{S^1\times S^2}}{\partial \dot \phi} = 2 T \int_{\theta_0}^\pi d\theta \frac{J_{12} \sin^2\theta - J_{11} \cos\theta + J_{10}}{\kappa \sqrt{2 \mu_1 \mu_2 (b_1^2 - k^4)} \sin\theta \sqrt{\cos\theta_0 - \cos\theta}},
\\
J_{12} &=& \kappa \mu_2 (k^4 - b_1^2),
\nonumber\\
J_{11} &=& (b_0 - b_1 + b_2 \cos\phi) (k^3 \mu_2^2 - k^3 \mu_1 \mu_2 - b_1 \kappa \mu_2) + b_1 \kappa (k \kappa - b_1 \mu_1 - b_1 \mu_2),
\nonumber\\
J_{10} &=& (b_0 - b_1 + b_2 \cos\phi) (k^3 \mu_2^2 - k^3 \mu_1 \mu_2  + b_1 \kappa \mu_1 - k \kappa^2) - b_1 \kappa (k \kappa - b_1 \mu_1 - b_1 \mu_2), \nonumber
\end{eqnarray}
and
\begin{eqnarray}
J_2 &=& 2 \int_{\theta_0}^\pi \, \frac{d\theta}{\theta'} \, \frac{\partial {\cal L}_{\rm NG}^{S^1\times S^2}}{\partial \dot \psi} = 2 T \int_{\theta_0}^\pi d\theta \frac{J_{22} \sin^2\theta - J_{21} \cos\theta + J_{20}}{\kappa \sqrt{2 \mu_1 \mu_2 (b_1^2 - k^4)} \sin\theta \sqrt{\cos\theta_0 - \cos\theta}},
\\
J_{22} &=& \kappa \mu_1 (k^4 - b_1^2),
\nonumber\\
J_{21} &=& (b_0 - b_1 + b_2 \cos\phi) (k^3 \mu_1^2 - k^3 \mu_1 \mu_2 - b_1 \kappa \mu_1) + b_1 \kappa (k \kappa - b_1 \mu_1 - b_1 \mu_2),
\nonumber\\
J_{20} &=& (b_0 - b_1 + b_2 \cos\phi) (k^3 \mu_1^2 - k^3 \mu_1 \mu_2  + b_1 \kappa \mu_2 - k \kappa^2) - b_1 \kappa (k \kappa - b_1 \mu_1 - b_1 \mu_2). \nonumber
\end{eqnarray}
For the 3-sphere case: $k = 1/2, b_0 = - b_1 = b/4, b_2 = 0$, the coefficients reduce to 
\begin{eqnarray}
&& J_{12} = \frac{\kappa \mu_2 (1 - b^2)}{16}, \qquad J_{22} = \frac{\kappa \mu_1 (1 - b^2)}{16},
\nonumber\\
&& J_{11} = J_{10} = \frac{b (- 2 \kappa^2 + \mu_2^2 - \mu_1 \mu_2 + b \kappa \mu_2 - b \kappa \mu_1)}{16},
\nonumber\\
&& J_{21} = J_{20} = \frac{b (- 2 \kappa^2 + \mu_1^2 - \mu_1 \mu_2 + b \kappa \mu_1 - b \kappa \mu_2)}{16},
\end{eqnarray}
and the angular momenta are
\begin{eqnarray}
J_1 &=& \frac{8 T}{\kappa \sqrt{2 \mu_1 \mu_2 (b^2 - 1)}} \left( 2 J_{12} \sqrt{\cos\theta_0 + 1} + \frac{\sqrt2 J_{10}}{\cos\frac{\theta_0}2} \ln \frac{2 \cos\frac{\theta_0}2}{\cos\frac{\pi}2} \right),
\\
J_2 &=& \frac{8 T}{\kappa \sqrt{2 \mu_1 \mu_2 (b^2 - 1)}} \left( 2 J_{22} \sqrt{\cos\theta_0 + 1} + \frac{\sqrt2 J_{20}}{\cos\frac{\theta_0}2} \ln \frac{2 \cos\frac{\theta_0}2}{\cos\frac{\pi}2} \right).
\end{eqnarray}
Moreover, the following combinations of angular momenta 
\begin{eqnarray}
J_- = J_1 - J_2
&=& - \frac{2 T \nu_1 (1 - b^2) \cos\frac{\theta_0}2}{\sqrt{(\nu_1^2 - \nu_2^2)(1 - b^2)}} - \frac{2 T b \nu_1 (\nu_2 + b \kappa)}{\kappa \sqrt{(\nu_1^2 - \nu_2^2) (1 - b^2)} \cos\frac{\theta_0}2} \ln \frac{2 \cos\frac{\theta_0}2}{\cos\frac{\pi}2},
\\
J_+ = J_1 + J_2
&=& \frac{2 T \nu_2 (1 - b^2) \cos\frac{\theta_0}2}{\sqrt{(\nu_1^2 - \nu_2^2)(1 - b^2)}} + \frac{2 T b (\nu_1^2 - \kappa^2)}{\kappa \sqrt{(\nu_1^2 - \nu_2^2) (1 - b^2)} \cos\frac{\theta_0}2} \ln \frac{2 \cos\frac{\theta_0}2}{\cos\frac{\pi}2},
\end{eqnarray}
are consistent with (\ref{J1SpikyS3}) and (\ref{J2SpikyS3}) respectively, as we can expect them from the coordinate transformation (\ref{hopf}).

\subsection{Giant magnon}
The second choice of parameters, $\kappa = k (\mu_1 - \mu_2), C_2 = C_1 k - b_1 \mu_2$, gives giant magnon which simplifies $\theta'$ and $\psi'$ as
\begin{eqnarray}
\theta' &=& \frac{(\mu_1 - \mu_2) \sqrt{2 \mu_1 \mu_2 (b_1^2 - k^4)} \sin\theta \sqrt{\cos\theta_1 - \cos\theta}}{\mu_1 (C_1 k - b_1 \mu_1 + b_2 \mu_2) \cos\theta - \mu_2 (C_1 k + b_1 \mu_1 - b_1 \mu_2)},
\\
\psi' &=& \frac{\mu_2 (C_1 k + b_1 \mu_1 - b_1 \mu_2) \cos\theta - \mu_1 (C_1 k - b_1 \mu_1 + b_1 \mu_2)}{\mu_1 (C_1 k - b_1 \mu_1 + b_1 \mu_2) \cos\theta - \mu_2 (C_1 k + b_1 \mu_1 - b_1 \mu_2)},
\end{eqnarray}
where
\begin{equation}
\cos\theta_1 = 1 - \frac{(C_1 k - b_1 \mu_1 - b_1 \mu_2)^2}{2 \mu_1 \mu_2 (b_1^2 - k^4)}.
\end{equation}
The energy of the magnon reads
\begin{eqnarray}
E &=& - 2 \int_{\theta_1}^\pi \, \frac{d\theta}{\theta'} \, \frac{\partial {\cal L}_{\rm NG}^{S^1\times S^2}}{\partial \dot t} 
= - 2 T \frac{\sqrt2 k [C_1^2 - k^2 (\mu_1 - \mu_2)^2)]}{(\mu_1 - \mu_2) \sqrt{2 \mu_1 \mu_2 (b_1^2 - k^4)} \cos\frac{\theta_1}2} \ln \frac{2 \cos\frac{\theta_1}2}{\cos\frac{\pi}2}.
\end{eqnarray}
For the particular values: $b_1 = -b/4, k^2 = 1/4, \mu_1 = \nu_2 + \nu_1, \mu_2 = \nu_2 - \nu_1$, then we have
\begin{eqnarray}
&&\sin\frac{\theta_1}2 = \sqrt{\frac{1 - \cos\theta_1}2} = \frac{C_1 + b \nu_2}{\sqrt{(\nu_1^2 - \nu_2^2)(1 - b^2)}}, \nonumber\\
&&E = \frac{2 T (\nu_1^2 - C_1^2)}{\nu_1 \sqrt{(\nu_1^2 - \nu_2^2) (1 - b^2)} \cos\frac{\theta_1}2} \ln\frac{2 \cos\frac{\theta_1}2}{\cos\frac{\pi}2},
\end{eqnarray}
which is identical with (\ref{EMagnonS3}).

The angular momenta again include contribution from $b_0$ and $b_2$. Their expressions are
\begin{eqnarray}
J_1 &=& 2 \int_{\theta_1}^\pi \, \frac{d\theta}{\theta'} \, \frac{\partial {\cal L}_{\rm NG}^{S^1\times S^2}}{\partial \dot \phi} = 2 T \int_{\theta_1}^\pi d\theta \frac{J_{12} \sin^2\theta + J_{11} \cos\theta - J_{10}}{(\mu_1 - \mu_2) \sqrt{2 \mu_1 \mu_2 (b_1^2 - k^4)} \sin\theta \sqrt{\cos\theta_1 - \cos\theta}},
\\
J_{12} &=& \mu_2 (\mu_1 - \mu_2) (k^4 - b_1^2),
\nonumber\\
J_{11} &=& C_1^2 k^2 - k^4 (\mu_1 - \mu_2)^2 - C_1 k (b_1 \mu_1 - b_0 \mu_2) - b_1 (\mu_2 - \mu_1) (b_1 \mu_1 + b_0 \mu_2)
\nonumber\\
&& + b_2 \mu_2 \cos\phi (C_1 k + b_1 \mu_1 - b_1 \mu_2),
\nonumber\\
J_{10} &=& C_1^2 k^2 - k^4 (\mu_1 - \mu_2)^2 + C_1 k (b_0 \mu_1 - b_1 \mu_2) + b_1 (\mu_2 - \mu_1) (b_0 \mu_1 + b_1 \mu_2)
\nonumber\\
&& + b_2 \mu_1 \cos\phi (C_1 k - b_1 \mu_1 + b_1 \mu_2). \nonumber
\end{eqnarray}
and
\begin{eqnarray}
J_2 &=& 2 \int_{\theta_1}^\pi \, \frac{d\theta}{\theta'} \, \frac{\partial {\cal L}_{\rm NG}^{S^1\times S^2}}{\partial \dot \psi} = 2 T \int_{\theta_1}^\pi d\theta \frac{J_{22} \sin^2\theta - J_{21} \cos\theta + J_{20}}{(\mu_1 - \mu_2) \sqrt{2 \mu_1 \mu_2 (b_1^2 - k^4)} \sin\theta \sqrt{\cos\theta_1 - \cos\theta}},
\\
J_{22} &=& \mu_1 (\mu_1 - \mu_2) (k^4 - b_1^2),
\nonumber\\
J_{21} &=& C_1^2 k^2 - k^4 (\mu_1 - \mu_2)^2 + C_1 k (b_0 \mu_1 - b_1 \mu_2) + b_1 (\mu_2 - \mu_1) (b_0 \mu_1 + b_1 \mu_2)
\nonumber\\
&& + b_2 \mu_1 \cos\phi (C_1 k + b_1 \mu_2 - b_1 \mu_1),
\nonumber\\
J_{20} &=& C_1^2 k^2 - k^4 (\mu_1 - \mu_2)^2 - C_1 k (b_1 \mu_1 - b_0 \mu_2) - b_1 (\mu_2 - \mu_1) (b_1 \mu_1 + b_0 \mu_2)
\nonumber\\
&& + b_2 \mu_2 \cos\phi (C_1 k + b_1 \mu_1 - b_1 \mu_2). \nonumber
\end{eqnarray}
For the 3-sphere case: $k = 1/2, b_0 = - b_1 = b/4, b_2 = 0$, we have
\begin{eqnarray}
&& J_{12} = \frac{\mu_2 (\mu_1 - \mu_2) (1 - b^2)}{16}, \qquad J_{22} = \frac{\mu_1 (\mu_1 - \mu_2) (1 - b^2)}{16},
\nonumber\\
&& J_{11} = J_{10} = J_{21} = J_{20} = \frac{C_1^2}4 + \frac{C_1 b (\mu_1 + \mu_2)}8 - \frac{(\mu_1 - \mu_2)^2 (1 - b^2)}{16},
\end{eqnarray}
and the angular momenta become
\begin{eqnarray}
J_1 &=& \frac{8 T}{(\mu_1 - \mu_2) \sqrt{2 \mu_1 \mu_2 (b^2 - 1)}} \left( 2 J_{12} \sqrt{\cos\theta_1 + 1} - \frac{\sqrt2 J_{10}}{\cos\frac{\theta_1}2} \ln \frac{2 \cos\frac{\theta_1}2}{\cos\frac{\pi}2} \right),
\\
J_2 &=& \frac{8 T}{(\mu_1 - \mu_2) \sqrt{2 \mu_1 \mu_2 (b^2 - 1)}} \left( 2 J_{22} \sqrt{\cos\theta_1 + 1} + \frac{\sqrt2 J_{20}}{\cos\frac{\theta_1}2} \ln \frac{2 \cos\frac{\theta_1}2}{\cos\frac{\pi}2} \right).
\end{eqnarray}
Similarly, the following combinations of angular momenta
\begin{eqnarray}
J_- &=& J_1 - J_2
= - \frac{2 T \nu_1 (1 - b^2) \cos\frac{\theta_1}2}{\sqrt{(\nu_1^2 - \nu_2^2)(1 - b^2)}} - \frac{2 T [C_1^2 + C_1 b \nu_2 - \nu_1^2 (1 - b^2)]}{\nu_1 \sqrt{(\nu_1^2 - \nu_2^2) (1 - b^2)} \cos\frac{\theta_1}2} \ln \frac{2 \cos\frac{\theta_1}2}{\cos\frac{\pi}2},
\\
J_+ &=& J_1 + J_2
= \frac{2 T \nu_2 (1 - b^2) \cos\frac{\theta_1}2}{\sqrt{(\nu_1^2 - \nu_2^2)(1 - b^2)}},
\end{eqnarray}
are identical to (\ref{J1MagnonS3}) and (\ref{J2MagnonS3}) respectively.

\section{Relation to sine-Gordon model}
In order to make connection to the sine-Gordon model \cite{Pohlmeyer:1975nb}, it is convenient to use the Polyakov action
\begin{eqnarray}
{\cal L}_{\rm P} &=& \frac{T}2 \Bigl\{ - (\partial_\tau t)^2 + (\partial_\sigma t)^2 + (\partial_\tau \vartheta)^2 - (\partial_\sigma \vartheta)^2 + \sin^2\vartheta \left[ (\partial_\tau \varphi_1)^2 - (\partial_\sigma \varphi_1)^2 \right]
\nonumber\\
&& \qquad + \cos^2\vartheta \left[ (\partial_\tau \varphi_2)^2 - (\partial_\sigma \varphi_2)^2 \right] + 2 b \sin^2\vartheta (\partial_\tau \varphi_1 \partial_\sigma \varphi_2 - \partial_\sigma \varphi_1 \partial_\tau \varphi_2) \Bigr\},
\end{eqnarray}
supplemented by the Virasoro constraint:
\begin{eqnarray}
- \partial_\tau t \partial_\sigma t + \partial_\tau \vartheta \partial_\sigma \vartheta + \sin^2\vartheta \partial_\tau \varphi_1 \partial_\sigma \varphi_1 + \cos^2\vartheta \partial_\tau \varphi_2 \partial_\sigma \varphi_2 &=& 0,
\\
- (\partial_\tau t)^2 - (\partial_\sigma t)^2 + (\partial_\tau \vartheta)^2 + (\partial_\sigma \vartheta)^2 + \sin^2\vartheta [ (\partial_\tau \varphi_1)^2 + (\partial_\sigma \varphi_1)^2 ] &&
\nonumber\\
+ \cos^2\vartheta [ (\partial_\tau \varphi_2)^2 + (\partial_\sigma \varphi_2)^2 ] &=& 0.
\end{eqnarray}
Now we will work with a similar, but slightly different embedding ansatz:
\begin{equation}
t = \kappa \tau, \quad \vartheta = \vartheta(y), \quad \varphi_1 = \nu_1 \tau + \psi_1(y), \quad \varphi_2 = \nu_2 \tau + \psi_2(y), \quad y = \alpha \sigma + \beta \tau.
\end{equation}
From the Virasoro constraint, one obtains the relation
\begin{equation} \label{VC1}
\beta \kappa^2 + A \nu_1 + B \nu_2 = 0,
\end{equation}
and the equation of motion
\begin{eqnarray}
\vartheta'^2 &=& \frac1{(\alpha^2-\beta^2)^2} \Bigl[ (\alpha^2 + \beta^2) \kappa^2 + 2 \alpha A b \nu_2 - \alpha^2 (\nu_1^2 + b^2 \nu_2^2) \sin^2\vartheta - \alpha^2 \nu_2^2 \cos^2\vartheta
\nonumber\\
&& \qquad - \frac{A^2}{\sin^2\vartheta} - \frac{(B + \alpha b \nu_1 \sin^2\vartheta)^2}{\cos^2\vartheta} \Bigr].
\end{eqnarray}
The rest of equations of motion read
\begin{eqnarray}
\psi_1' &=& \frac1{\alpha^2 - \beta^2} \left( \frac{A}{\sin^2\vartheta} + \beta \nu_1 - \alpha b \nu_2 \right),
\\
\psi_2' &=& \frac1{\alpha^2 - \beta^2} \left( \frac{B + \alpha b \nu_1 \sin^2\vartheta}{\cos^2\vartheta} + \beta \nu_2 \right).
\end{eqnarray}
In order to have desired boundary condition, i.e. $\vartheta' = 0$ at $\vartheta = \pi/2$, one should impose the following constraints on the parameters
\begin{equation}
(\alpha^2 + \beta^2) \kappa^2 + 2 \alpha A  b \nu_2 - \alpha^2 (\nu_1^2 + b^2 \nu_2^2) - A^2 = 0, \qquad B + \alpha b \nu_1 = 0,
\end{equation}
which, together with the constraint (\ref{VC1}), leads to two possible solutions
\begin{eqnarray}
&&\beta \kappa = - \alpha \nu_1, \quad A = \alpha (\kappa + b \nu_2),\quad B = - \alpha b \nu_1;
\nonumber\\
&&\qquad \kappa = \nu_1, \quad A = - \beta \nu_1 + \alpha b \nu_2 , \quad B = - \alpha b \nu_1.
\end{eqnarray}
corresponding to spiky string and giant magnon respectively.  The value of $B$ in both cases is chosen to avoid divergence of $\psi_2'$ at the equator.

We also explicit the evaluation of worldsheet action for later use,
\begin{equation}
\sqrt{-h} = \frac{\alpha^2}{\alpha^2 - \beta^2} (\kappa^2 - \nu_1^2 \sin^2\vartheta - \nu_2^2 \cos^2\vartheta).
\end{equation}

\subsection{Spiky String}
With the choice of $\beta \kappa = - \alpha \nu_1, A = \alpha (\kappa + b \nu_2), B = - \alpha b \nu_1$ and a turning point at  $\vartheta = \pi/2$, we claim to have a spiky solution:
\begin{eqnarray}
\vartheta' &=& \frac{\alpha \sqrt{(\nu_1^2 - \nu_2^2)(1 - b^2)} \, \cos\vartheta}{(\alpha^2 - \beta^2) \, \sin\vartheta} \sqrt{\sin^2\vartheta - \sin^2\vartheta_0},
\nonumber\\
\sin\vartheta_0 &\equiv& \frac{- \alpha \nu_1 + \beta b \nu_2}{\beta \sqrt{(\nu_1^2 - \nu_2^2) (1 - b^2)}}.
\end{eqnarray}
We may find solution of $\vartheta$,
\begin{equation}
\cos\vartheta = \frac{\cos\vartheta_0}{\cosh(D_0 y)}, \qquad D_0 \equiv \frac{\alpha \sqrt{(\nu_1^2 - \nu_2^2)(1 - b^2)}}{\alpha^2 - \beta^2} \cos\vartheta_0.
\end{equation}
The worldsheet action reduces to
\begin{equation}
\sqrt{-h} = \frac{\alpha^2 \nu_1^2}{\beta^2} + \frac{\alpha^2 (\nu_1^2 - \nu_2^2)}{\alpha^2 - \beta^2} \cos^2\vartheta = \kappa^2 + \frac{(\alpha^2 - \beta^2) D_0^2}{1 - b^2} \frac1{\cosh^2(D_0 y)}.
\end{equation}
If we impose $(\alpha^2 - \beta^2) D_0^2 = - \kappa^2 (1 - b^2)$ and define the field
\begin{equation}
\sin^2\Psi_0 = - \frac{(1 - b^2) \sqrt{-h}}{(\alpha^2 - \beta^2) D_0^2} = \tanh^2(D_0 y),
\end{equation}
then we recover the sine-Gordon equation,
\begin{equation}\label{sG_spike}
\partial_\tau^2 \Psi_0 - \partial_\sigma^2 \Psi_0 - \frac{(\alpha^2 - \beta^2) D_0^2}2 \sin{(2 \Psi_0)} = 0.
\end{equation}

The conserved quantities can be computed by the following formula
\begin{equation}
E
= - 2 \int_{\vartheta_0}^{\frac{\pi}2} \, \frac{d\vartheta}{\alpha \vartheta'} \, \frac{\partial {\cal L}_{\rm P}}{\partial \dot t}, \qquad
J_1 = 2 \int_{\vartheta_0}^{\frac{\pi}2} \, \frac{d\vartheta}{\alpha \vartheta'} \, \frac{\partial {\cal L}_{\rm P}}{\partial \dot \varphi_1}, \qquad
J_2 = 2 \int_{\vartheta_0}^{\frac{\pi}2} \, \frac{d\vartheta}{\alpha \vartheta'} \, \frac{\partial {\cal L}_{\rm NG}}{\partial \dot \varphi_2},
\end{equation}
which give exactly the same results as in NG formulation.

Now we are ready to identify a spiky string as a kink or soliton of mass parameter $m_b \equiv \kappa \sqrt{1 - b^2}$ and velocity $v \equiv \beta / \alpha > 1$ in the corresponding sine-Gordon equation (\ref{sG_spike}), provided
\begin{equation}
\tan\left( \frac{\Psi_0}2 - \frac{\pi}4 \right) =  \mathrm{e}^{\pm D_0 y} = \mathrm{e}^{\pm m_b \gamma_v (\sigma + v \tau)}, \qquad \gamma_v \equiv (v^2 - 1)^{-1/2},
\end{equation}
where a plus (minus) sign is assigned to (anti-)kink solution.

\subsection{Giant Magnon}
With the choice of $\kappa = \nu_1, A =  - \beta \nu_1 + \alpha b \nu_2, B = - \alpha b \nu_1$ and a turning point at $\vartheta = \pi/2$, we claim to have a magnon solution:
\begin{eqnarray}
\vartheta' &=& \frac{\alpha \sqrt{(\nu_1^2 - \nu_2^2)(1 - b^2)} \, \cos\vartheta}{(\alpha^2 - \beta^2) \, \sin\vartheta} \sqrt{\sin^2\vartheta - \sin^2\vartheta_1},
\nonumber\\
\sin\vartheta_1 &\equiv& \frac{\beta \nu_1 - \alpha b \nu_2}{\alpha \sqrt{(\nu_1^2 - \nu_2^2) (1 - b^2)}}.
\end{eqnarray}
We may find solution of $\vartheta$,
\begin{equation}
\cos\vartheta = \frac{\cos\vartheta_1}{\cosh(D_1 y)}, \qquad D_1 \equiv \frac{\alpha \sqrt{(\nu_1^2 - \nu_2^2)(1 - b^2)}}{\alpha^2 - \beta^2} \cos\vartheta_1.
\end{equation}
The worldsheet action reduces to
\begin{equation}
\sqrt{-h} = \frac{\alpha^2 (\nu_1^2 - \nu_2^2)}{\alpha^2 - \beta^2} \cos^2\vartheta = \frac{(\alpha^2 - \beta^2) D_1^2}{1 - b^2} \frac1{\cosh^2(D_1 y)}.
\end{equation}
Therefore, if we define the field
\begin{equation}
\sin^2\Psi_1 = \frac{(1 - b^2) \sqrt{-h}}{(\alpha^2 - \beta^2) D_1^2} = \frac1{\cosh^2(D_1 y)},
\end{equation}
then the sine-Gordon equation is recovered,
\begin{equation}\label{sG_magnon}
\partial_\tau^2 \Psi_1 - \partial_\sigma^2 \Psi_1 + \frac{(\alpha^2 - \beta^2) D_1^2}2 \sin(2 \Psi_1) = 0.
\end{equation}
Once again, we will have the same physical quantities with $C_1 = - \beta \nu_1 / \alpha$.

Now we are ready to identify a magnon as a kink of mass parameter $m_b \equiv \kappa \sqrt{1 - b^2}$ and velocity $v \equiv \beta / \alpha < 1$ in the corresponding sine-Gordon equation (\ref{sG_magnon}), provided
\begin{equation}
\tan{\frac{\Psi_1}{2}} = \mathrm{e}^{\pm D_1 y} = \mathrm{e}^{\pm m_b \gamma_v (\sigma + v \tau)}, \qquad \gamma_v \equiv (1 - v^2)^{-1/2}
\end{equation}
where a plus (minus) sign is assigned to (anti-)kink solution.  Solutions of N (anti-)kinks can be easily constructed and correspond to configuration of multiple (anti-)magnons.  In particular, a solution corresponds to scattering between two magnons, in the center of mass frame, reads
\begin{equation}
\tan\frac{\Psi_1}2 = u \frac{\sinh(m_b \gamma_u u \tau)}{\cosh(m_b \gamma_u \sigma)}, \qquad \gamma_u \equiv (1 - u^2)^{-1/2},
\end{equation}
and scattering between magnon and anti-magnon is
\begin{equation}
\tan\frac{\Psi_1}2 = \frac1{u} \frac{\sinh(m_b \gamma_u u \tau)}{\cosh(m_b \gamma_u \sigma)}, \qquad \gamma_u \equiv (1 - u^2)^{-1/2}.
\end{equation}
Moreover, a breather solution (bound state of magnon and anti-magnon) is
\begin{equation}
\tan\frac{\Psi_1}2 = \frac1{a} \frac{\sin(m_b \gamma_a a \tau)}{\cosh(m_b \gamma_a \sigma)}, \qquad \gamma_a \equiv (1 + a^2)^{-1/2}.
\end{equation}

\section{Fractional spin chain from $B$-deformed SYM}
If we carefully inspect the conserved quantities and dispersion relation, we may find that a simple redefinition,
\begin{equation}\label{b-relation}
\tilde{\lambda} = (1 - b^2) \lambda, \qquad \tilde{E} = (1 - b^2) E,
\end{equation}
can bring us back to the ordinary description without $B$-field.  That is\footnote{Here we would like to take the convention $\tilde{J}_i = J_i, C_1 = 0$ in the giant magnon solution.},
\begin{eqnarray}
\label{dispersion_spike}  \tilde{J}_1 = \sqrt{\tilde{J}_2^2 + 4 \tilde{T}^2 \cos^2{\vartheta_0}} \qquad && \text{for spiky string},
\\
\label{dispersion_magnon} \tilde{E} - \tilde{J}_1 = \sqrt{\tilde{J}_2^2 + \frac{\tilde{\lambda}}{\pi^2} \sin^2{\frac{p}{2}}} \qquad && \text{for giant magnon}.
\end{eqnarray}
Therefore, one is tempted to claim a dual super Yang-Mills theory\footnote{The SYM in our mind would be $D = 4, {\cal N} = 4$ if $S^3$ is part of $S^5$ or $D = 2, {\cal N} = (2, 2)$ if we are dealing with NS1-NS5 system.} deformed by such a $B$-field, denoting  $B$-deformed SYM, can be realized by a simple replacement (\ref{b-relation}).  In particular, the spin-chain description in the $SU(2)$ sector of $B$-deformed SYM consists of operators of the form,
\begin{equation}
{\cal O} \sim \mathrm{Tr}[\Phi_1^{\tilde{J}_1} \Phi_2^{\tilde{J}_2}] + \cdots,
\end{equation}
where $\Phi_1$ and $\Phi_2$ are two complex adjoint scalars of the theory and the dots denote all possible orderings of the fields.  With no deformation, $J_1$ and $J_2$ are two integer-valued charges corresponding to a $U(1)\times U(1)$ subgroup of the R-symmetry group.  With deformation, they are rescaled by a factor $\sqrt{1 - b^2}$ and no need to be integers.  Therefore we obtain a {\it fractional} chain of non-integer length $\tilde{L} = \tilde{J}_1 + \tilde{J}_2$.  To be precise, at one-loop the dilatation operator in the $SU(2)$ sector can be mapped onto the Hamiltonian of the Heisenberg spin chain \cite{Minahan:2002ve}.  The scaling dimension $\Delta$ is related to the energy $M$
\begin{equation}\label{spinchain}
\Delta = \tilde{L} + \frac{\lambda_e}{8 \pi^2} M, \qquad M = \sum_k{\varepsilon(p_k)} = \sum_k{4\sin^2{(\frac{p_k}{2})}},
\end{equation}
where each magnon carries individual energy $\varepsilon(p_k)$ and momentum $p_k$.  At the limit $\tilde{\lambda} \ll \tilde{J}_2^2$, equation (\ref{dispersion_magnon}) agrees with (\ref{spinchain}), up to ${\cal O}(\lambda_e)$, for single magnon if we identify
\begin{equation}
\Delta = \tilde{E}, \qquad \lambda_e = \frac{\tilde{\lambda}}{\tilde{J}_2^2}.
\end{equation}
This provides a strong evidence to support our conjecture of existence of a $B$-deformed SYM dual to our gravity background.

\section{Discussion}
We have studied a rotating string on $S^3$ with background NS-NS $B$-field and found solutions of spiky string and giant magnon as two limiting cases.  The dispersion relation between regularized conserved charges is still in a simple expression but refined by (\ref{b-relation}).  It is interesting to compare our results with those from other deformed backgrounds with nontrivial NS-NS $B$-field.  In the Lunin-Maldacena background, for instance, $\beta$-deformation causes a shift in the magnon momentum \cite{Chu:2006ae}, and the dual field theory is the ${\cal N} = 1$ SYM, which is marginally deformed from ${\cal N} = 4$.  In our background, $B$-field has the effect of recaling string tension, and we have suspect a $B$-deformed SYM to be the dual description, justified by matching the dispersion relation on both sides.

We also succeeded in mapping our solutions into the kink solutions in the associated sine-Gordon model.  With the replacement of mass parameter $m_b$, one is ready to study scattering as in \cite{Hofman:2006xt}.

As a future direction, it might be interesting to investigate the finite size effect to our solutions \cite{Arutyunov:2006gs, Lee:2008ui}, where one expects to set one of turning points to be located at some other $\theta$ instead of the equator. Moreover, by rewriting $S^3$ as a Hopf fibration, we are able to carry out our results to the study of rotating string on other backgrounds such as deformed $AdS_3\times S^3$ \cite{Israel:2004vv} or $AdS_4\times CP^3$ \cite{Aharony:2008ug,Ahn:2008ua}.  This work is in progress.

\acknowledgments
The authors would like thank Hsien-Chung Kao, Pei-Ming Ho and Takeo Inami for useful discussion.
The authors are supported by the Taiwan's National Science Council and National Center for Theoretical Sciences under Grant No. NSC 96-2112-M-008-006-MY3 (CMC, JHT), NSC97-2112-M-002-015-MY3 (WYW) and NSC97-2119-M-002-001.


\end{document}